\documentstyle[12pt,psfig]{article}
\newcommand{\beq}{\begin{equation}}
\newcommand{\eeq}{\end{equation}}
\newcommand{\beqa}{\begin{eqnarray}}
\newcommand{\eeqa}{\end{eqnarray}}
\newcommand{\ba}{\begin{array}}
\newcommand{\ea}{\end{array}}
\newcommand{\CR}{\nonumber \\}

\newcommand{\A}{\alpha}

\renewcommand{\thefootnote}{\fnsymbol{footnote}}

\newcommand{\nin}{\;\,/\!\!\!\!\!\in}

\makeatletter
\@addtoreset{equation}{section}

\makeatother

\begin{document}

\begin{titlepage}
\null
\begin{flushright}
YITP-00-26
\\
UT-893
\\
hep-th/0005164
\\
May, 2000
\end{flushright}

\vskip 1.5cm
\begin{center}

  {\LARGE String Junctions in $B$ Field Background}

\lineskip .75em
\vskip 2cm
\normalsize

  {\large Jiro Hashiba$^1$, Kazuo Hosomichi$^1$ and Seiji Terashima$^2$}

\vskip 1.5cm

  {\large \it $^1$Yukawa Institute for Theoretical Physics,\\
               Kyoto University, Kyoto 606-8502, Japan}

\vskip 0.5cm

  {\large \it $^2$Department of Physics, University of Tokyo,\\
               Tokyo 113-0033, Japan}

\vskip 2cm

{\bf Abstract}
\end{center}

It has been recently shown that F-theory on $K3$ with background $B$
 fields (NSNS and RR 2-forms) is dual to the CHL string in 8
 dimensions. In this paper, we reexamine this duality in terms of string 
 junctions in type IIB string theory. It is in particular stressed that
 certain 7-brane configurations produce $Sp$ gauge groups in a novel way.

\end{titlepage}

\renewcommand{\thefootnote}{\arabic{footnote}}
\baselineskip=0.7cm

\clearpage

%%%%%%%%%%%%%%%%%%%%%%%%%%%%%%%%%%%%%%%%%%%%%%%%%%%%%%%%%%%%%%%%%%%%%%%%%%%%%
%%%%%%%%%%%%%%%%%%%%%%%%%%%%%%%%%%%%%%%%%%%%%%%%%%%%%%%%%%%%%%%%%%%%%%%%%%%%%
\section{Introduction}
%%%%%%%%%%%%%%%%%%%%%%%%%%%%%%%%%%%%%%%%%%%%%%%%%%%%%%%%%%%%%%%%%%%%%%%%%%%%%
%%%%%%%%%%%%%%%%%%%%%%%%%%%%%%%%%%%%%%%%%%%%%%%%%%%%%%%%%%%%%%%%%%%%%%%%%%%%%

In recent developments in string theory, the notion of duality has
played a crucial role in the understanding of its nonperturbative
properties. Some of the dualities are clearly understood in terms of
twelve dimensional F-theory \cite{Vafa,MoVa}. The ``definition'' of
F-theory is as follows. We consider F-theory compactified on a
Calabi-Yau $(n+1)$-fold $X_{n+1}$ which admits an elliptic fibration
$\pi : X_{n+1} \rightarrow B_n$, where $B_n$ is an $n$ complex
dimensional manifold. Then, this theory is dual to type IIB string
theory compactified on $B_n$ with 7-branes wrapped on divisors in
$B_n$. Via this type IIB description, various string vacua can be
analyzed in the context of F-theory.

The power of F-theory would be well illustrated by the duality between
F-theory and heterotic string. For instance, F-theory compactified on an
elliptically fibered $K3$ surface is dual to heterotic string
compactified on a 2-torus. In this duality, the Narain moduli space on
the heterotic side, which contains the k\"{a}hler and complex moduli of
the 2-torus as well as the Wilson line data, coincides with the complex
structure moduli space of the $K3$ on the F-theory side. The origin of
gauge symmetry in 8 dimensions is quite different in these dual
theories. In heterotic theory, gauge symmetry comes from the breaking of
the $E_8 \times E_8$ or $SO(32)$ gauge group to its subgroups. On the
other hand, F-theory acquires unbroken gauge symmetry when the $K3$
surface develops the corresponding $ADE$ type singularity.

The definition of F-theory implies that the 8 dimensional theory
given above is represented also by type IIB string theory on $B_1={\bf
P}^1$ with 24 7-branes located at points on the ${\bf P}^1$. The
condition that the $K3$ surface in the F-theory description develops a
singularity, namely the condition that a gauge symmetry enhancement
occurs, corresponds to the situation where in type IIB theory
several 7-branes coalesce at the same position. The perturbative spectra
of heterotic string are identified with string junctions stretched
between 7-branes. The 7-brane configurations which give rise to $ADE$
type gauge groups have been obtained in \cite{DeZw,Jo,GaZw}. 

By modding out $E_8 \times E_8$ heterotic string compactified on a circle 
by some involution, we can construct another heterotic theory which is now
referred to as the CHL string \cite{ChHoLy,ChPo}. It has been recently
found by Bershadsky, Pantev and Sadov that the CHL string in 8
dimensions also has a dual description in terms of F-theory on a
$K3$ \cite{BePaSa}. The point is that incorporating $B$ field background
in the F-theory part is equivalent to the CHL involution. Specifically, we
switch on NSNS and/or RR 2-form fluxes along ${\bf P}^1$ so that the
$SL(2,{\bf Z})$ symmetry of type IIB string is broken to its subgroup
$\Gamma_0(2)$ \cite{BePaSa,BeKlMaTh}. Then, the complex moduli space of
the $K3$ is confined to its subspace, which is identical with the moduli
space for the CHL string.

It should be noted that $Sp$ gauge groups are realized in the CHL
string in 8 dimensions and its F-theory counterpart. This is a
remarkable feature absent in the ordinary vacua where no flux is turned
on.
\footnote{
It has also been found in \cite{Witten,Im} that $Sp$ gauge groups
appear when several $D7$-branes meet an $O7^+$-plane.
}
Our aim in this paper is to analyze string junctions in the presence
of non-vanishing $B$ fields, and to clarify how $Sp$ gauge groups are
realized in the type IIB framework.

The paper is organized as follows. In section 2, we give a brief review
of the CHL string in 8 dimensions. In section 3, the F-theory dual
description of the CHL string is reviewed. The CHL involution is
realized in the F-theory context by the presence of quantized $B$
fluxes. We consider type IIB string theory in section 4, and translate
the geometric data of F-theory into string junctions. The last section is 
devoted to conclusions and discussions.

%%%%%%%%%%%%%%%%%%%%%%%%%%%%%%%%%%%%%%%%%%%%%%%%%%%%%%%%%%%%%%%%%%%%%%%%%%%%%
%%%%%%%%%%%%%%%%%%%%%%%%%%%%%%%%%%%%%%%%%%%%%%%%%%%%%%%%%%%%%%%%%%%%%%%%%%%%%
\section{Review of The CHL String}
%%%%%%%%%%%%%%%%%%%%%%%%%%%%%%%%%%%%%%%%%%%%%%%%%%%%%%%%%%%%%%%%%%%%%%%%%%%%%
%%%%%%%%%%%%%%%%%%%%%%%%%%%%%%%%%%%%%%%%%%%%%%%%%%%%%%%%%%%%%%%%%%%%%%%%%%%%%

   Here we give a review of the eight-dimensional CHL string theory
 and its possible gauge symmetry enhancements.
 In the following argument we take
 the left-moving sector to be supersymmetric,
 and we use the bosonic realization for the right-moving sector.

   Let us begin with the ordinary toroidal compactification.
 In the $T^d$-compactification of the $E_8\times E_8$ heterotic string
 we have $d$ periodic coordinates
 $X^i(z,\bar{z}) = X^i(z)+ \tilde{X}^i(\bar{z})$
 and sixteen antiholomorphic bosons $X^I(\bar{z})$.
 Their mode expansions take the following form\footnote
{ Here and throughout the section we use the convention $\alpha'=2$.}
\begin{eqnarray}
  X^i(z) &=&
    x^i - ip^i\ln z + i\sum_n \frac{\alpha^i_n}{nz^n}, \nonumber \\
  \tilde{X}^i(\bar{z}) &=&
    \tilde{x}^i - i\tilde{p}^i\ln \bar{z}
     + i\sum_n \frac{\tilde{\alpha}^i_n}{n\bar{z}^n}, \\
  \tilde{X}^I(\bar{z}) &=&
    \tilde{x}^I - i\tilde{p}^I\ln \bar{z}
     + i\sum_n \frac{\tilde{\alpha}^I_n}{n\bar{z}^n},  \nonumber
\end{eqnarray}
 and the canonical quantization leads to
 the following commutation relations:
\begin{equation}
\begin{array}{ccccl}
  \left[ x^i , p_j \right] &\!\!=\!\!&
  \left[ \tilde{x}^i , \tilde{p}_j \right] &\!\!=\!\!&
  i\delta^i_{~j}, \\
  \left[ \alpha_m^i,\alpha_n^j \right] &\!\!=\!\!&
  \left[ \tilde{\alpha}_m^i, \tilde{\alpha}_n^j \right] &\!\!=\!\!&
  m g^{ij}\delta_{m+n,0},
\end{array}
~~
\begin{array}{ccl}
  \left[ \tilde{x}^I, \tilde{x}^J \right] &\!\!=\!\!&
  i\delta^{IJ}, \\
  \left[ \tilde{\alpha}_m^I, \tilde{\alpha}_n^J\right] &\!\!=\!\!&
  m\delta^{IJ}\delta_{m+n,0}.
\end{array}
\end{equation}
 The momenta $(p_i,\tilde{p}_i, \tilde{p}_I)$
 are quantized in the following way:
\begin{eqnarray}
  \tilde{p}_I &=& q_I + A_{iI}w^i, \nonumber \\
  \tilde{p}_i &=& n_i + (-g_{ij}-B_{ij}+A_i^IA_j^I)\frac{w^j}{2} + A_{iI}q^I, 
   \label{hetero1} \\
          p_i &=& n_i + ( g_{ij}-B_{ij}+A_i^IA_j^I)\frac{w^j}{2} + A_{iI}q^I,
  \nonumber
\end{eqnarray}
\begin{equation}
 n_i,w^i\in {\bf Z}~,~~~
 (q^I)\equiv \vec{q}=(\vec{q}_1,\vec{q}_2)\in \Gamma_8\oplus\Gamma_8.
 \label{hetero2}
\end{equation}
 Here the (half-)integers $(n_i, w^i, q^I)$ correspond respectively to
 the momenta, the winding numbers and the gauge charges,
 and specify a point in the Narain lattice $\Gamma^{d,d+16}$.
 The background fields $g_{ij},B_{ij},A_{iI}$ parameterize
 the following moduli space of vacua:
\[
  O(d,d+16,\Gamma^{d,d+16})\backslash O(d,d+16)/(O(d)\times O(d+16)) ~.
\]
 In fact, if we define the norm of the momentum vector by
$ p^2 \equiv p_ip^i -\tilde{p}_i\tilde{p}^i -\tilde{p}^I\tilde{p}^I$,
 it is indeed independent of the moduli:
\[
 p^2 \equiv
       p_ip^i -\tilde{p}_i\tilde{p}^i -\tilde{p}^I\tilde{p}^I
     = 2n_iw^i - q^Iq^I
     \in 2{\bf Z} ~.
\]
   A choice of a vacuum uniquely determines
 the decomposition of $V\equiv {\bf R}^{d,d+16}$
 into two subspaces $V_\pm$,
 each having positive and negative-definite norm.
 The gauge symmetry enhancements occur precisely
 when some lattice vectors of squared norm $-2$
 are contained in $V_-$.
 Thus the analysis of possible gauge enhancements
 is reduced to a purely mathematical problem.

~

   The CHL string is defined as the asymmetric ${\bf Z}_2$ orbifold
 of the $E_8\times E_8$ heterotic string theory compactified onto
 torus.
 The ${\bf Z}_2$ symmetry to be considered is
 the exchange of two $E_8$'s accompanied by a half-period shift.

   The orbifolding gives rise to two sectors.
 In the untwisted sector we have the same quantization of the momenta
 (\ref{hetero1}), (\ref{hetero2})
 as in the usual toroidal compactification,
 and we only take the ${\bf Z}_2$-symmetric states as physical states.
 In the twisted sector we have
\[
 (\vec{n},\vec{w})\in ({\bf Z}^d,{\bf Z}^d+\vec{v}) ~,
\]
 where $\vec{v}$ denotes the half-period shift vector.
 Namely if $\vec{v}$ is taken along the first direction we have
 $w^1\in {\bf Z}+\frac{1}{2}$.
 The quantization of the sixteen antiholomorphic bosons gives
\[
 \vec{q}=(\vec{q}_1,\vec{q}_1)~~ (2\vec{q}_1\in \Gamma_8)
  ~,~~~~~
 \tilde{\alpha}^I_m = (-)^{2m}\tilde{\alpha}^{I\pm 8}_m
  ~~(m\in {\bf Z}/2).
\]

   In order for the orbifolding to be possible,
 the Wilson line must be invariant under the exchange
 of two $E_8$'s, namely $ A_i^I = A_i^{I\pm 8} $.
 Hence the moduli space is reduced to the following form:
\begin{equation}
 O(d,d+8,\Gamma_{(d)})\backslash O(d,d+8)/(O(d)\times O(d+8))
   ~,~~~
 \Gamma_{(d)} \equiv  \Gamma_8\oplus \Gamma^{d-1,d-1}(2)\oplus \Gamma^{1,1}.
\end{equation}
\[
  \Gamma^{d,d}(2)\equiv\left\{\vec{x}=(\xi_i,\eta^i)\in {\bf Z}^{2d} \right\}
  ~,~~
  \vec{x}^2= 4\xi_i\eta^i.
\]
   Here the lattice $\Gamma_{(d)}$ is the Narain lattice
 of the CHL string \cite{Mik},
 and any automorphism of this lattice is a symmetry
 of the CHL string theory.
 The way to associate a point of $\Gamma_{(d)}$
 to a state with charges $(n_i,w^i,\vec{q})$
 is given as follows:
\[
 {\bf x}\equiv
 (n_i,2w^i,\vec{Q}\equiv \vec{q}_1+\vec{q}_2) \in \Gamma_{(d)}
 ~,~~~
 {\bf x}^2\equiv 4n_iw^i - \vec{Q}^2.
\]
 As in the case of ordinary toroidal compactification,
 a choice of vacuum determines the decomposition of any vector ${\bf x}$
 into left and right-moving projection components, 
 ${\bf x} = {\bf x}_+ + {\bf x}_-$.

   We define a classification of lattice points
 $\left\{{\bf x}\right\}$ as follows: \\ \hspace*{2cm}
   type 1.~~ ${\bf x}\in 2(\Gamma^{1,1}\oplus
               \Gamma_8)\oplus \Gamma^{d-1,d-1}(2)$.  \\ \hspace*{2cm}
   type 2.~~ ${\bf x}$ is not in the above subset,
             and ${\bf x}^2$ is divisible by 4. \\ \hspace*{2cm}
   type 3.~~ ${\bf x}^2$ is not divisible by 4. \\
 In the table below we summarize the classification
 of string states according to the above definition.
 Here the half-period shift is chosen in the first direction.
%\begin{table}[hbt]
\begin{center}
\begin{tabular}{|c||l|l|l|c|}\hline
 (A)& 2$w^1$ even (untwisted sector)
      &$Q\in  2\Gamma_8$ & $n_1$   even & type 1 \\ \cline{1-1} \cline{4-5}
 (B)& &                  & $n_1$   odd  & type 2 \\ \cline{1-1} \cline{3-5}
 (C)& &$Q\nin 2\Gamma_8$ & $Q^2/2$ even & type 2 \\ \cline{1-1} \cline{4-5}
 (D)& &                  & $Q^2/2$ odd  & type 3 \\ \hline
 (E)& \multicolumn{2}{|l|}{$2w^1$ odd (twisted sector)}
      &      $n_1+Q^2/2$ even           & type 2 \\ \cline{1-1} \cline{4-5}
 (F)& \multicolumn{2}{|l|}{}
      &      $n_1+Q^2/2$ odd            & type 3 \\ \hline
\end{tabular}
\end{center}
%\end{table}

   A careful evaluation of the mass-shell condition
 yields the following mass formula:
 the mass of the lightest state among those corresponding to
 a given vector ${\bf x}$ of type $i$ is given by 
\begin{equation}
  m^2 =  \frac{{\bf x}_+^2}{2} = -\frac{{\bf x}_-^2}{2}-N_i ~,~~~
  (N_i=2,0,1 ~~~\mbox{for  }i= 1,2,3).
\end{equation}
 Here $N_i$ are obtained as the sum of the zero-point energy and
 the term $\sim (\vec{q}_1-\vec{q}_2)^2$.
 Remarkably the mass formula depends only on the type of
 ${\bf x}$ and not on any of the additional informations.
 Hence the massless charged vector bosons arise either from
 the ${\bf x}$ of type 1 and squared norm $-4$,
 or of type 3 and squared norm $-2$.

~

   Let us focus on the $T^2$ compactification,
 and take the half-period along the first direction.
 To study the possible enhancements of gauge symmetry,
 it is the best way to consider first the charged vector bosons
 with ${\bf x}^2=-4$, corresponding to long roots.
 From the previous argument it follows that
 all the long roots ${\bf x}$ arise from the type 1
 of the three subsets.
 Further analysis leads us to the following lemma: \\
{\it
(1)  Any two long roots are orthogonal. \\
(2)  The average of any two long roots belongs to $\Gamma_{(2)}$. \\
}
 These can be proved straightforwardly once we realize that
 $n_2, w^2$ must be odd for any long roots.
 From these lemma it follows that,
 given the set of long roots ${\bf x}_{j~(j=1,\cdots, 2+n)}$,
 the lattice vectors
\[
 \left\{ \pm{\bf x}_j ~,~~~ (\pm{\bf x}_i\pm'{\bf x}_j)/2 \right\}
\]
 form the root system of $Sp(2+n)$.
 Hence the CHL string can realize the $Sp$ type gauge symmetry
 in 8 dimensions.

   The set of long roots for $Sp(2+n)$ can be explicitly
 constructed in the following way\cite{ChPo}.
 Take an $A_n$ sublattice of $\Gamma_8$ and choose its basis
 $e_{i~ (i=1,\cdots n)}$ so that they satisfy
 $ e_i\cdot e_j = 1+\delta_{ij}$.
 Then the long roots are expressed as
\begin{equation}
\begin{array}{lclrrrr}
               & &  n_1 &w^1    &n_2    &w^2    &\vec{Q}    \\
  {\bf x}_1    &=& (0,  &\!\! 0,&\!\! 1,&\!\!-1,&\!\! 0   ), \\
  {\bf x}_2    &=& (2,  &\!\!-1,&\!\! 1,&\!\! 1,&\!\! 0   ), \\
  {\bf x}_{2+j}&=& (0,  &\!\!-1,&\!\! 1,&\!\! 1,&\!\! 2e_j).
\end{array}
\end{equation}

   Generically the enhanced gauge symmetry is given by
 the product of the $Sp$ type level one algebra and
 the $ADE$ type level two algebra.
 For example, using the formula
 $  A_n\times E_{8-n} \subset E_8 $
 we find that the following enhancements are possible:
\begin{equation}
 Sp(2+n)\times E_{8-n} ~~~(n=0,\cdots, 8),
\label{CHL1}
\end{equation}
where we have used the notation $E_5=SO(10),E_4=SU(5),E_3=SU(3) \times
SU(2),E_2=SU(2) \times U(1)$ and $E_1=SU(2)$.

%%%%%%%%%%%%%%%%%%%%%%%%%%%%%%%%%%%%%%%%%%%%%%%%%%%%%%%%%%%%%%%%%%%%%%%%%%%%%
%%%%%%%%%%%%%%%%%%%%%%%%%%%%%%%%%%%%%%%%%%%%%%%%%%%%%%%%%%%%%%%%%%%%%%%%%%%%%
\section{F-theory Dual of The CHL String}
%%%%%%%%%%%%%%%%%%%%%%%%%%%%%%%%%%%%%%%%%%%%%%%%%%%%%%%%%%%%%%%%%%%%%%%%%%%%%
%%%%%%%%%%%%%%%%%%%%%%%%%%%%%%%%%%%%%%%%%%%%%%%%%%%%%%%%%%%%%%%%%%%%%%%%%%%%%

In this section we present the F-theory compactification to 8
dimensions which precisely yields a dual description of the CHL string
explained in the previous section, following
\cite{BePaSa,BeKlMaTh}. Since the ordinary Narain moduli space is
restricted to its subspace due to the CHL involution, we have to find a
similar mechanism which freezes some of the complex moduli of $K3$ so as
to obtain the same moduli space. This is systematically done by turning
on 2-form background fields $B$ taking values in $H^2({\bf P}^1,{\bf
Z}/2)$.

Let us consider the case where 2-forms $B_{NS}$ and $B_R$ take the
values
\begin{equation}
  \label{flux}
  \left(\begin{array}{c}
          B_{NS} \\
          B_R
        \end{array}\right) =
  \left(\begin{array}{c}
          1/2 \\
          0
        \end{array}\right).
\label{bf}
\end{equation}
If $(B_{NS},B_R)$ is non-vanishing, the $SL(2,{\bf Z})$ monodromy group
around 7-branes on ${\bf P}^1$ must be reduced to a smaller group
which keeps $(B_{NS},B_R)$ invariant modulo $H^2({\bf P}^1,{\bf Z})$. In 
the present case (\ref{flux}), the reduced monodromy group is
$\Gamma_0(2)$, which is represented by the matrices of the form
\begin{equation}
  \Gamma_0(2) = \left\{
                \left(\begin{array}{cc}
                        a & b \\
                        c & d
                      \end{array}\right) ; a,b,d \in {\bf Z}, c \in 2{\bf Z},
                                           ad-bc=1 \right\}.
\label{Gamma0(2)}
\end{equation}
The elliptic $K3$ surface with a section can be expressed by the
familiar Weierstrass equation, which in general has the entire monodromy
group $SL(2,{\bf Z})$, not $\Gamma_0(2)$. In order to get the elliptic
$K3$ with $\Gamma_0(2)$ monodromy, which we denote by $X$, the
Weierstrass equation must take a special form. To find the form of the
equation, note that the elliptic fibration of the $K3$ must have one
more section which intersects the fiber at the point corresponding to a
half-period of the fiber torus. We thus conclude that $X$ is of the form
\begin{equation}
  \label{gamma02}
    X~:~ y^2 = (x-a_4(z))(x^2+a_4(z)x+b_8(z)),
\end{equation}
where $z$ is the coordinate on the ${\bf P}^1$ base, and $a_4(z),b_8(z)$
are polynomials of degree 4 and 8, respectively. The Weierstrass
form (\ref{gamma02}) indeed has a section $s_1:(x=a_4(z),y=0)$, in
addition to the ordinary one $s_0:(x=\infty,y=\infty)$. If we denote by
$\alpha$ and $\beta$ the two independent 1-cycles of the fiber torus
which correspond to NSNS and RR directions, then the additional section
$s_1$ corresponds to the half-period $(1/2)\alpha + 0\cdot\beta$, in
agreement with the quantized flux (\ref{flux}).

The discriminant $\Delta$ of $X$ is given by
\begin{equation}
  \label{discriminant}
  \Delta = (4b_8-a_4^2)(b_8+2a_4^2)^2.
\end{equation}
This expression shows that $X$ always has eight $A_1$ type singularities,
and eight $I_1$ type singular fibers. All the eight $A_1$ singularities
are situated at the section $s_1$.

The $K3$ surface $X$ given by (\ref{gamma02}) with $\Gamma_0(2)$
monodromy can be constructed from another $K3$ surface $Y$, through the
following two steps. First, we mod out $Y$ by an involution $\sigma$, to 
make the ``intermediate'' $K3$ $\tilde{X}=Y/\sigma$. The involution
$\sigma$ is none other than a geometric realization of the CHL
involution. Second, we apply a birational transformation to $\tilde{X}$,
yielding the desired $K3$ surface $X$. Let us start with the
construction of $Y$. We take the $K3$ surface $Y$ to be a double cover
of the Hirzebruch surface ${\bf F}_0 = {\bf P}^1 \times {\bf P}^1$,
which branches along the zero locus of a section of the line bundle ${\cal
O}_{{\bf F}_0}(4,4)$ over ${\bf F}_0$. Denoting the inhomogeneous
coordinates on the two ${\bf P}^1$ factors in ${\bf F}_0$ by $s$ and
$t$, we can represent $Y$ as the hypersurface equation
\begin{equation}
  \label{K3Y}
    Y ~:~ y^2 = \sum_{i,j=0}^4 a_{i,j} s^i t^j.
\end{equation}
Here, $a_{i,j}$ parametrize the complex structure moduli of $Y$. By
fixing either one of $s$ and $t$, one can see that $Y$ admits two
elliptic fibrations $p:Y \rightarrow {\bf P}^1$ and $q:Y \rightarrow {\bf
P}^1$ whose bases are the two ${\bf P}^1$ factors in ${\bf F}_0$
parametrized by $s$ and $t$, respectively.

Let us turn to the first step, the application of the involution
$\sigma$ to $Y$. The involution $\sigma$ is defined by the ${\bf Z}_2$
action on $Y$,
\begin{equation}
  \label{K3-involution}
  \sigma~:~(y,s,t) \rightarrow (-y,s,-t).
\end{equation}
This means that $\sigma$ inverses the orientation of one of the
${\bf P}^1$ and the elliptic fiber over it simultaneously. Since $Y$
should be invariant under the ${\bf Z}_2$ action (\ref{K3-involution})
in order to be modded out by $\sigma$, $Y$ has to take the form
\begin{equation}
  \label{K3Y2}
  y^2 = \sum_{i=0}^4 \sum_{j=0}^2 a_{i,2j} s^i t^{2j}.
\end{equation}
The hypersurface (\ref{K3Y2}) is embedded in ${\bf C}^3$ parametrized by 
$(y,s,t)$ which is reduced to ${\bf C}^2/{\bf Z}_2 \times {\bf C}$ by
the involution $\sigma$. The orbifold ${\bf C}^2/{\bf Z}_2$ is described 
by the three well-defined coordinates $(u=y^2,v=t^2,w=yt)$ and the
relation among them
\begin{equation}
  \label{orbifold}
  uv=w^2.
\end{equation}
By combining (\ref{K3Y2}) with (\ref{orbifold}), we can express
$\tilde{X}=Y/\sigma$ by the hypersurface equation parametrized by
$(w,s,v)$,
\begin{eqnarray}
  \label{tildeX}
  \tilde{X} ~:~ w^2 &=& v P(s,v), \CR
             P(s,v) &=& \sum_{i=0}^4 \sum_{j=0}^2 a_{i,2j} s^i v^{j}.
\end{eqnarray}
Note that, similarly to $Y$, $\tilde{X}$ can also be interpreted as a
double cover of ${\bf F}_0$, which is now coordinatized by $s$ and
$v$. The only difference between $Y$ and $\tilde{X}$ is that the branch
divisor defining $\tilde{X}$ consists of three components
$\tilde{C}_1:\{v=\infty\}$, $\tilde{C}_2:\{v=0\}$ and
$\tilde{C}_3:\{P(s,v)=0\}$, whereas the branch divisor for a generic $Y$
consists of a single component. In other words, the involution $\sigma$
has the effect of restricting the complex moduli space of $Y$, so that
the branch divisor defining $Y$ splits to $\tilde{C}_1$, $\tilde{C}_2$
and $\tilde{C}_3$, which are identified with the zero loci of sections
of line bundles $q^*{\cal O}_{{\bf P}^1}(1)$, $q^*{\cal O}_{{\bf
P}^1}(1)$ and ${\cal O}_{{\bf F}_0}(4,2)$.

The second step is to convert $\tilde{X}$ into $X$ by a birational
transformation. To do this, we focus on the fact that $\tilde{X}$ has
eight $A_1$ type singularities, four of which are on $\tilde{C}_1$, and
the other four on $\tilde{C}_2$. The four $A_1$ singularities on
$\tilde{C}_1$ comes from the four points where $\tilde{C}_3$ intersects
$\tilde{C}_1$. The same is true for the singularities on
$\tilde{C}_2$. By blowing up ${\bf F}_0$ at the four points on
$\tilde{C}_1$, we obtain a surface birational to ${\bf F}_0$,
whose generic ${\bf P}^1$ fiber becomes reducible at the four points on
the ${\bf P}^1$ base where $\tilde{C}_3$ meets $\tilde{C}_1$. Each of
the four reducible fibers consists of two ${\bf P}^1$'s, one of which is
generated by the blowing up procedure. Then, we blow down the remaining
four ${\bf P}^1$'s in the reducible fibers. The resulting surface is
${\bf F}_4$, on which there exist three divisors $C_i~(i=1,2,3)$ which
inherit from $\tilde{C}_{i}$. From the above birational transformation,
it is easy to see that $C_2$ intersects $C_3$ at eight points, while
$C_1$ does not intersect $C_3$ at all. Finally, we identify $X$ with the 
double cover of ${\bf F}_4$, whose branch divisor consists of
$C_i~(i=1,2,3)$. The two sections $s_0$ and $s_1$ of $X$ originate from
$C_1$ and $C_2$, and $X$ indeed has eight $A_1$ singularities on $s_1$
which appear from the eight points where $C_2$ and $C_3$ intersect. It
is also important to note that the elliptic fibration $\pi$
for $X$ traces back to the elliptic fibration $p$ for $Y$.

The $K3$ surface $X$ constructed in this way can be represented by the
equation (\ref{gamma02}), which has monodromy group $\Gamma_0(2)$. Then
we switch on the background flux (\ref{flux}), to ensure that the
monodromy group of $X$ is not enlarged to $SL(2,{\bf Z})$.

To determine the unbroken gauge symmetry in F-theory on $X$, we need to
examine the relation between the singularity structures of $Y$ and
$X$. In what follows, we shall not make an exhaustive analysis, but
illustrate the main idea by just one example. The case to be considered
in this paper is an $A_{2n-1}$ type singularity in $Y$, which is located
on a fixed elliptic fiber $\{t=0\}$ with respect to the fibration $q$. By
appropriately adjusting the coefficients $a_{i,2j}$ in (\ref{K3Y2}), we
can let $Y$ develop an $A_{2n-1}~(n=1,\ldots,8)$ singularity at the point
$\{s=t=0\}$. Explicitly, the $K3$ surface $Y$ with such a singularity is 
given by
\begin{eqnarray}
  \label{Ysingularity}
     A_1 &:& y^2 = t^4 + t^2 + s^2, \CR
     A_3 &:& y^2 = t^4 + st^2 + s^2, \CR
     A_5 &:& y^2 = st^4 + t^4 + 2st^2 + s^2, \CR
     A_7 &:& y^2 = s^2t^4 + t^4 + 2st^2 + s^2, \CR
     A_9 &:& y^2 = s^3t^4 + t^4 + 2st^2 + s^2, \CR
  A_{11} &:& y^2 = s^4t^4 + t^4 + 2st^2 + s^2, \CR
  A_{13} &:& y^2 = s^4t^4+s^3t^4+2s^2t^4+t^4+s^4t^2+2s^3t^2+2st^2+s^2, \CR
  A_{15} &:& y^2 = s^3t^4 + t^4 + s^4t^2 + 2st^2 + s^2.
\end{eqnarray}
If we resolve the above $A_{2n-1}$ singularity,
there appear, from the singularity, $2n-1$ ${\bf P}^1$'s which intersect
one another according to the $A_{2n-1}$ Dynkin diagram. Let us denote
these $2n-1$ components by $S_i~(i=1,\ldots,2n-1)$. Under the involution
$\sigma$, the middle component $S_n$ remains unchanged, whereas the
other $2n-2$ components are exchanged in the manner
\begin{equation}
  \label{automorphism}
  S_i \leftrightarrow S_{2n-i}~~~(i=1,\ldots,n-1).
\end{equation}
This suggests that $\sigma$ acts on $S_i$'s as an outer automorphism of
the $A_{2n-1}$ algebra, giving rise to the new algebra $C_n$.

Therefore we expect that $Sp(n)$ gauge symmetry enhancement takes place
in F-theory on $X$, when $Y$ has an $A_{2n-1}$ type singularity on
$\{t=0\}$ (or on $\{t=\infty\}$). The next problem is then to determine
what type of singularity $X$ should develop in order for $Sp(n)$ gauge
symmetry to be produced. This is easily done by just translating the
singularity data (\ref{Ysingularity}) for $Y$ into the one for
$X$. It turns out that $\tilde{X}$ develops an
$A_3$ (resp. $D_{n+2}~(n=2,\ldots,8)$) singularity when $Y$ has a
singularity of the type $A_1$ ({\rm resp.} $A_{2n-1}$). The explicit
form of $\tilde{X}$ can be determined by using (\ref{K3Y2}),
(\ref{tildeX}) and (\ref{Ysingularity}) to be
\begin{eqnarray}
     A_3 &:& w^2 = v(v^2 + v + s^2), \CR
     D_4 &:& w^2 = v(v^2 + sv + s^2), \CR
     D_5 &:& w^2 = v(sv^2 + v^2 + 2sv + s^2), \CR
     D_6 &:& w^2 = v(s^2v^2 + v^2 + 2sv + s^2), \CR
     D_7 &:& w^2 = v(s^3v^2 + v^2 + 2sv + s^2), \CR
     D_8 &:& w^2 = v(s^4v^2 + v^2 + 2sv + s^2), \CR
     D_9 &:& w^2 = v(s^4v^2+s^3v^2+2s^2v^2+v^2+s^4v+2s^3v+2sv+s^2), \CR
  D_{10} &:& w^2 = v(s^3v^2 + v^2 + s^4v + 2sv + s^2).
\label{ds}
\end{eqnarray}
In the present situation, the singularity structures of $X$ and
$\tilde{X}$ are the same. We are thus led to the following dictionary
relating the type of singularity in $X$ to gauge symmetry:
\begin{equation}
  \begin{array}{c|c}
    {\rm singularity~in}~ X & {\rm gauge~symmetry} \\ \hline
                        A_3 & Sp(1) \\
     D_{n+2}~(n=2,\ldots,8) & Sp(n)
  \end{array}
\end{equation}

One might wonder why $A_3$ (resp. $D_{n+2}$) singularity does not give rise to
$SU(4)$ (resp. $SO(2n+4)$), but to $Sp$ gauge symmetry. This is due to
the presence of quantized fluxes along the base of $X$. Recall that $X$
necessarily has eight $A_1$ singularities and eight $I_1$ fibers. The
$A_1$ singularities cannot be deformed, since deforming them would force 
$X$ to have the full monodromy group $SL(2,{\bf Z})$, and therefore conflict
with the presence of background fluxes. The $A_3$ singularity in $X$ is
made from the collision of two $A_1$ singularities. Similarly, the
$D_{n+2}$ singularity in $X$ comes from the collision of two $A_1$
singularities and $n$ $I_1$ fibers. If the $A_1$ singularities could be
deformed, then $SU(4)$ or $SO(2n+4)$ gauge symmetry would be generated
as it should be. However, since deforming the $A_1$ singularities is in
fact forbidden, it is possible that $Sp$ gauge groups are realized
though it seems strange at first sight. In the next section, we will
find further evidence that $Sp(n)~(n=1,\ldots,8)$ gauge symmetry is
actually generated, by explicitly constructing the string junctions
which realize the simple roots for $Sp(n)$ Lie algebra.

%%%%%%%%%%%%%%%%%%%%%%%%%%%%%%%%%%%%%%%%%%%%%%%%%%%%%%%%%%%%%%%%%%%%%%%%%%%%%
%%%%%%%%%%%%%%%%%%%%%%%%%%%%%%%%%%%%%%%%%%%%%%%%%%%%%%%%%%%%%%%%%%%%%%%%%%%%%
\section{String Junction}
%%%%%%%%%%%%%%%%%%%%%%%%%%%%%%%%%%%%%%%%%%%%%%%%%%%%%%%%%%%%%%%%%%%%%%%%%%%%%
%%%%%%%%%%%%%%%%%%%%%%%%%%%%%%%%%%%%%%%%%%%%%%%%%%%%%%%%%%%%%%%%%%%%%%%%%%%%%

In this section we consider 
the type IIB string theory compactified on $B_1={\bf P}^1$.
There are 24 7-branes located at points on the ${\bf P}^1$,
and we turn on the background $B$ field as explained in
the previous section.
The nonzero $B$ fields do not break the supersymmetry,
because the supersymmetry transformation law of fermions
contain only their field strengths $H=dB$, which simply vanish
in this case.
We will focus on the vicinity of the 7-branes realizing the
$Sp$ type gauge symmetry.
Indeed, such an enhancement of the gauge symmetry is possible
in the CHL string theory, and we may as well expect the same symmetry
also in the F-theory by constructing the $K3$ surface $X$
from the other $K3$ surface $Y$ by an orbifolding.
However, it is not clear whether we can realize $Sp(n)$
gauge symmetry by the above construction of $X$, since
it is only $X$ and not $Y$ which is of physical significance.
This is true even when we consider the M-theory and membranes
by further compactification on $S^1$.
In the following we explain that $Sp(n)~(n \geq 2)$ gauge symmetry
can be realized by an $SO(2n+4)$-type fiber singularity,
by using string junctions and taking account of the
background $B$ field appropriately. The case of $Sp(1) \simeq SU(2)$
will be discussed at the end of this section.

We follow the notation of \cite{DeHaIqZw} and in particular
denote by $X_{[p,q]}$ the $[p,q]$ 7-brane
or the 7-brane on which $[p,q]$-string can end. 
We also denote the 7-branes 
$X_{[1,0]}, X_{[1,-1]},X_{[1,1]}$ and $X_{[0,1]}$
as ${\bf A}, {\bf B},{\bf C}$ and ${\bf D}$, respectively, 
for notational simplicity.
The monodromy of a $[p,q]$ 7-brane is given by the formula 
\beq
K_{[p,q]} = \left(
\begin{array}{cc}
      1+pq    &  -p^2    \\
      q^2 &  1-pq
    \end{array}
\right),
\eeq
and the 7-branes ${\bf A}, {\bf B},{\bf C}$ and ${\bf D}$
have the following monodromy matrices
\beq
K_{\bf A} = \left(
\begin{array}{cc}
      1    &  -1    \\
      0 &  1
    \end{array}
\right),
K_{\bf B} = \left(
\begin{array}{cc}
      0    &  -1    \\
      1 &  2
    \end{array}
\right),
K_{\bf C} = \left(
\begin{array}{cc}
      2    &  -1    \\
      1 &  0
    \end{array}
\right),
K_{\bf D} = \left(
\begin{array}{cc}
      1    &  0    \\
      1 &  1
    \end{array}
\right).
\eeq

The $D_{n+2}~(n \geq 2)$ singularity of $X$ (\ref{ds}),
which is necessary for realizing the $Sp(n)$ gauge symmetry,
corresponds to the collection of 7-branes ${\bf A}^{n+2} {\bf B} {\bf C}$
in the type IIB picture. 
The $D_{n+2}$ singularity can be deformed to two $A_1$ singularities
and $n$ $I_1$ fibers.
However, further resolution is not allowed due to the presence of
the background $B$ field.
In order to see this in terms of 7-brane configurations,
we notice the equivalence between ${\bf A}^{n+2} {\bf B} {\bf C}$ 
and ${\bf A}^{n} {\bf D}^2 {\bf C}^2$.
This can be seen as follows.
First, by moving a 7-brane across the branch cut of another 7-brane
we obtain the following formulae \cite{GaHaZw}
\beq
X_{[p,q]} X_{[r,s]}=X_{[r,s]+(ps-qr) [p,q]} X_{[p,q]}
=X_{[r,s]} X_{[p,q]+(ps-qr)[r,s]},
\eeq
which give, in particular,
${\bf A} {\bf B}= {\bf D}{\bf A}, 
 {\bf A} {\bf D}={\bf D} {\bf C}$ and 
${\bf C} {\bf A}={\bf D} {\bf C}$.
Using them we obtain 
\beq
{\bf A}^{n+2} {\bf B} {\bf C}={\bf A}^{n+1} {\bf D} {\bf A} {\bf C}
={\bf A}^{n} {\bf D} {\bf C} {\bf A} {\bf C}
={\bf A}^{n} {\bf D}^2 {\bf C}^2.
\eeq

From the new expression for the collection of 7-branes
we identify the two $A_1$ singularities with
${\bf D}^2$ and ${\bf C}^2$,
and $I_1$ fibers with ${\bf A}$'s.
Indeed, ${K_{{\bf A}}}, {K_{{\bf D}}}^2$ and ${K_{{\bf C}}}^2$
are elements of $\Gamma_0(2)$ defined in (\ref{Gamma0(2)})
although ${K_{{\bf D}}}$ and ${K_{{\bf C}}}$ are not.
The VEV of the $n+2$ complex scalars, which are 
in the multiplets containing the gauge fields for
the Cartan subalgebra of the gauge symmetry,
represent the positions 
of the $n$ ${\bf A}$'s and
relative positions of the two ${\bf D}$'s and two ${\bf C}$'s on the
${\bf P}^1$. Each of these scalars originates from 
the open string whose two endpoints are on a single 7-brane.

The massless BPS states are represented by string junctions
which connect the 7-branes collapsing to a single point on ${\bf P}^1$.
If there is no background $B$ field,
the string junctions connecting some of the 7-branes
${\bf A}^{n+2} {\bf B}{\bf C}$ form the root lattice of
$SO(2 n+4)$ \cite{DeZw}.
We show in fig. $1$ the junctions which correspond to
the simple roots\footnote{
In this paper we use same symbol for a simple root and a junction
corresponding to it.}
$\A_i$ for $SO(2 n+4)$.

\begin{figure}
    \centerline{\psfig{figure=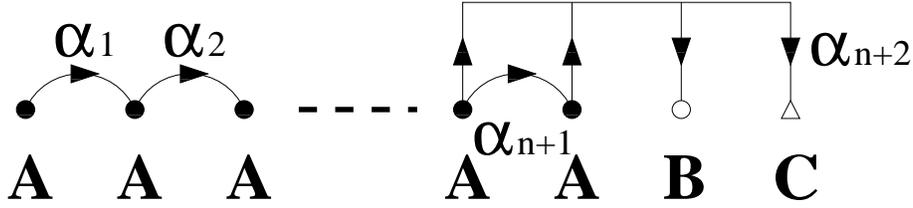,width=12cm}}
    \caption{The 7-brane configuration and junctions which
 realize $SO(2n+4)$.}
\end{figure}

%\vspace{1cm}
%fig. $1$ is the picture of 
%$A_1 A_2 \cdots A_i \cdots A_{n+1} A_{n+2} B  C$ and string junctions;
%\\ $\A_i (1 \leq i \leq n+1)$ : string connecting $A_i$  and $A_{i+1}$
%\\ $\A_{n+2}$ : string connecting $A_{n+1},A_{n+2}$  and $B,C$
%\vspace{1cm}

Now we analyze how the junctions are transformed
by the chain of rearrangements of 7-branes that brings
${\bf A}^{n+2} {\bf B} {\bf C}$ into ${\bf A}^{n} {\bf D}^2 {\bf C}^2$.
By the Hanany-Witten effect \cite{HaWi,DeZw}
a $[p,q]$-string and a $[r,s]$ 7-brane passing through each other
create $(ps-qr)$ $[r,s]$-strings between the two.
Hence the string junction which have $n_1$ endpoints on $X_{[p,q]}$
and $n_2$ on $X_{[r,s]}$
are transformed according to the following identity
\[
 n_1[p,q]+n_2[r,s] = n_2\left\{[r,s]+(ps-qr)[p,q]\right\}
                   + \left\{n_1-(ps-qr)n_2\right\}[p,q]
\]
in the process of the rearrangement
$X_{[p,q]}X_{[r,s]}=X_{[r,s]+(ps-qr)[p,q]}X_{[p,q]}$.
Using this formula we find that the junction corresponding to
the $\alpha_i$ in fig. 1 is transformed to that in fig. 2.

\begin{figure}
    \centerline{\psfig{figure=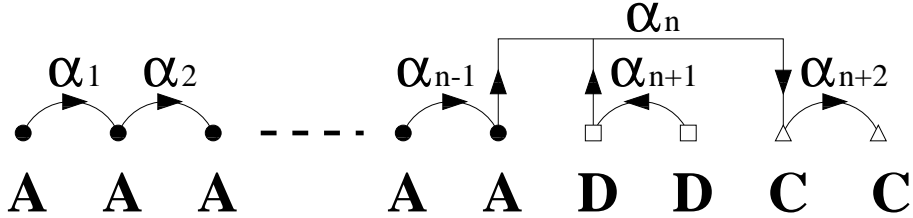,width=12cm}}
    \caption{Another representation of the 7-brane configuration and
 junctions for $SO(2n+4)$.}
\end{figure}

%\vspace{1cm}
%fig. $2$ is the picture of 
%$A_1 A_2 \cdots A_i \cdots A_{n-1} A_{n} D_1 D_2 C_1 C_2$ 
%and string junctions;
%\\ $\A_i (1 \leq i \leq n-1)$ : string connecting $A_i$  and $A_{i+1}$
%\\ $\A_{n}$ : string connecting $D_1,C_1$  and $A_{n}$
%\\ $\A_{n+1}$ : string connecting $D_2$  and $D_1$
%\\ $\A_{n+2}$ : string connecting $C_2$  and $C_1$
%\vspace{1cm}

Once we turn on the $B$ field,
the gauge fields corresponding to the junction
 $\A_{j}~ (n \! \leq \! j \! \leq \! n+2)$
disappear from the massless spectrum.
This is because the subgroup $SU(2)$ of the $SO(2 n+4)$  
which is generated by $\A_{j}$ and $-\A_{j}$ 
contains the fluctuation modes that separate the two
${\bf D}$'s or two ${\bf C}$'s.
In other words,
since the $A_1$ singularities cannot be deformed,
the scalars corresponding to their deformation parameters
cannot have nonzero VEV and must not remain massless.
On the other hand, junctions which have the same number of endpoints
on each of the two $\bf D$'s (and each of the two $\bf C$'s) remain
massless, since they correspond to the fluctuations
that move ${\bf D}^2$ (and ${\bf C}^2$) altogether
without breaking them into constituents.
By taking these rules into account, we conclude that the massless
spectrum representing the simple roots for $Sp(n)$ is generated by
  $\tilde{\A}_{i} \!\equiv\! \A_{i}~( 1 \! \leq \! i \! \leq \! n-1)$
 and
\beq
\tilde{\A}_{n} \equiv \A_{n} + (\A_{n} +\A_{n+1}+\A_{n+2}),
\label{longroot}
\eeq
as depicted in fig. 3.

\begin{figure}
    \centerline{\psfig{figure=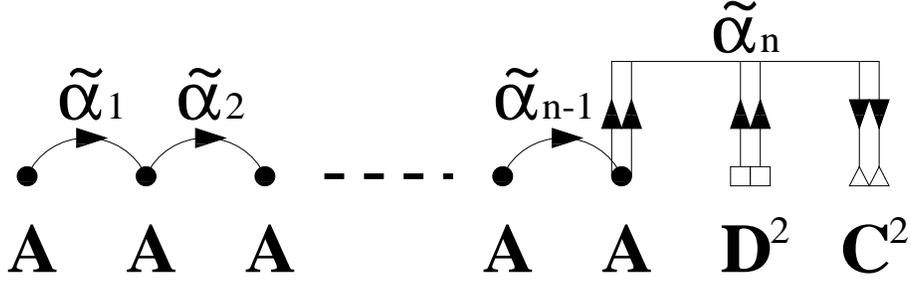,width=12cm}}
    \caption{The simple roots for $Sp(n)$.}
\end{figure}

In particular we can check that   
$\tilde{\A}_{n-1} \cdot \tilde{\A}_{n} = -2$ 
and that the root $\tilde{\A}_{n}$ has the norm
$(\tilde{\A}_{n})^2=4$, namely it is a long root.
Note that the inner product of the roots is given by
minus the intersection number of the corresponding
string junctions, i.e.
  ${\bf J}_{\A_i} \cdot {\bf J}_{\A_j} =-\A_i \cdot \A_j$
where ${\bf J}_{\A_i}$ is the junction corresponding to the root $\A_i$.
It is also known that,
by compactifying further onto $S^1$ and going to the M-theory picture,
the intersection number of junctions is identified with
that of membranes wrapping on the corresponding two-cycles in $K3$.
Furthermore we expect the BPS condition to be unchanged
when we turn on the background $B$ field \cite{MaMoMiSt}.\footnote{
The junctions corresponding to the frozen moduli are also
expected to be BPS, but they do not yield massless states
according to the argument in the previous paragraph.}
Thus we can realize $Sp(n)$ gauge group in type IIB string theory
by 7-branes and the non-vanishing background $B$ field.

Here we notice that, according to the arguments in \cite{DeHaIqZw, MiNeSe},
the BPS junction ${\bf J}$ have to satisfy
\beq
{\bf J} \cdot {\bf J} = 2g -2 +b \geq -2,
\eeq
where $g$ and $b$ are, the genus and the number of the boundary, 
of the corresponding two-cycles in $K3$.
Thus it seems impossible to have BPS junctions
with self-intersection number $-4$ corresponding
to the long roots for $Sp(n)$.
However, this is not true because every junction corresponding to 
the long roots consists of two junctions orthogonal to each other.
For example, $\tilde{\A}_n$ is expressed as the sum of
$\A_n$ and $(\A_{n} +\A_{n+1}+\A_{n+2})$,
which would satisfy the BPS condition if the background $B$ field were
turned off.
Moreover the two junctions are equivalent as cycles of
the corresponding $K3$, since under the presence of the $B$ field
the $A_1$ singularities cannot be deformed.
Thus every junction corresponding to a long root is a bound state
of two junctions which cannot be separated due to the background $B$ field.
Note that when we turn off the $B$ field, 
it becomes a marginal bound state and can be separated into two BPS junctions.

The 7-brane configuration for $Sp(1)$ has an expression slightly
different from the ones for the sequence $Sp(n)~(n \geq 2)$. The
configuration is given by two pairs of ${\bf D}$ branes, namely ${\bf
D}^2{\bf D}^2$. As pointed out before, this configuration would give
$SU(4)$ gauge symmetry if both pairs of ${\bf D}$ branes could be
separated. However, we have $Sp(1)$ symmetry here, since the rule
concerning massless spectrum allows only one string junction $\alpha$
consisting of two coincident D-strings, which emanate from one of the
two ${\bf D}^2$'s and get absorbed into the other as depicted in
fig. 4. We identify this junction with the simple root for $Sp(1)$.

\begin{figure}
    \centerline{\psfig{figure=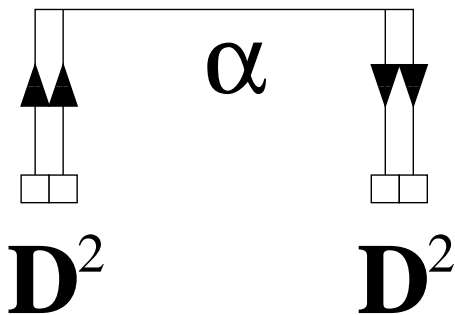,width=6cm}}
    \caption{The 7-brane configuration and simple root for $Sp(1)$.}
\end{figure}

The $O7^+$-plane has also been used to produce $Sp(n)$ gauge group,
and a certain dual description of CHL string theory 
contains it \cite{BeKlMaTh, Witten}.
In this description,
a modification of the BPS condition of string junction
is required \cite{Im}. 
In fact this is same as our result ${\bf J}{\bf J}=-4$ for the long
root.
The $O7^+$-plane might be realized in our picture by
eliminating the $B$ field by a singular gauge transformation
and localize it to a single point on ${\bf P}^1$.
By identifying the $O7^+$-plane with the point
onto which the nonzero $B$ field is localized,
we might be able to prove the equivalence between
the F-theory with quantized flux and the description with the
$O7^+$-plane.

%%%%%%%%%%%%%%%%%%%%%%%%%%%%%%%%%%%%%%%%%%%%%%%%%%%%%%%%%%%%%%%%%%%%%%%%%%%%%
%%%%%%%%%%%%%%%%%%%%%%%%%%%%%%%%%%%%%%%%%%%%%%%%%%%%%%%%%%%%%%%%%%%%%%%%%%%%%
\section{Conclusions and Discussions}
%%%%%%%%%%%%%%%%%%%%%%%%%%%%%%%%%%%%%%%%%%%%%%%%%%%%%%%%%%%%%%%%%%%%%%%%%%%%%
%%%%%%%%%%%%%%%%%%%%%%%%%%%%%%%%%%%%%%%%%%%%%%%%%%%%%%%%%%%%%%%%%%%%%%%%%%%%%

In this paper we have clarified how the $Sp$ type gauge symmetry
can be realized in type IIB string theory on ${\bf P}^1$
with quantized $B$ flux.
In the presence of the $B$ flux the monodromy around the 7-branes
are restricted to the subgroup $\Gamma_0(2)$ of $SL(2,{\bf Z})$,
and due to this condition some 7-branes are forced to be bound together.
The analysis of the string junctions on the collapsed 7-branes
shows that, by excluding the junctions corresponding to the deformation
that would enlarge the monodromy group, we can obtain $Sp(n)$ gauge group
out of the $D_{n+2}$ singular fiber.
Some junctions in the root system of $Sp(n)$ have the self-intersection
$-4$ and at first sight they cannot be BPS saturated.
However, a careful look into the corresponding cycles in $K3$
leads us to realize that the long roots always correspond to
two degenerate cycles, so that they are BPS saturated.

It would be a very interesting problem to see the full correspondence
between the CHL string and type IIB string theory with
nonzero $B$ field.
To see this, note that the gauge groups of the type (\ref{CHL1})
is possible in the CHL string.
However the singular fiber of the type $E_6$ or $E_8$ is impossible
in the type IIB string with the $B$ field, since the monodromies around
the corresponding collections of 7-branes are not in $\Gamma_0(2)$.
In \cite{BePaSa} the realization of such gauge groups as well as
$Sp(9), Sp(10)$ has been given in a very complicated way.
Some enhancements require that the monodromy group become
further smaller to the group $\Gamma(2)$.
It would be interesting to study how such gauge groups are realized
by 7-branes and string junctions in the type IIB string theory.

On the other hand, according to our construction of $Sp$ gauge symmetry
the gauge groups of the form $Sp \times Sp$ are apparently possible,
which is in contradiction with the result in the CHL string
and the description with the $O7^+$-plane.
In the CHL string we can easily see that one cannot realize the
product of $Sp$'s, since by the lemma (2) of section 2
the embedding of the root lattice of  $Sp\times Sp$ in $\Gamma_{(2)}$
would result in the embedding of the larger $Sp$ gauge group.
Translating this fact into our framework we must conclude that,
if we realize two $Sp$ singularity at two different points on ${\bf P}^1$
we obtain a large $Sp$ group, with some junctions linking two distant
fiber singularities becoming massless.
At present we have no idea for resolving this discrepancy.

%%%%%%%%%%%%%%%%%%%%%%%%%%%%%%%%%%%%%%%%%%%%%%%%%%%%%%%%%%%%%%%%%%%%%%%%%%%%%
%%%%%%%%%%%%%%%%%%%%%%%%%%%%%%%%%%%%%%%%%%%%%%%%%%%%%%%%%%%%%%%%%%%%%%%%%%%%%
\section*{Acknowledgements}
%%%%%%%%%%%%%%%%%%%%%%%%%%%%%%%%%%%%%%%%%%%%%%%%%%%%%%%%%%%%%%%%%%%%%%%%%%%%%
%%%%%%%%%%%%%%%%%%%%%%%%%%%%%%%%%%%%%%%%%%%%%%%%%%%%%%%%%%%%%%%%%%%%%%%%%%%%%

We would greatly like to thank Y.~Yoshida for collaboration at the early 
stage of this work. We also thank T.~Takayanagi for useful
discussions. J.~H. and K.~H. thank K.~Ohta and S.~Sugimoto for helpful
comments. The work of J.~H., K.~H. and S.~T. was supported in part
by the Japan Society for the Promotion of Science under the Postdoctoral
Research Programs No. 11-09295, No. 12-02721 and No. 11-08864, respectively.

\end{document}